\RequirePackage{amsmath} 

\documentclass[runningheads,a4paper]{llncs}

\usepackage{times}
\usepackage{amssymb}
\usepackage{graphicx}
\usepackage{url}

\usepackage[T1]{fontenc}

\usepackage{etoolbox}
\newtoggle{blindforreview}
\togglefalse{blindforreview}   

\usepackage{amsmath}
\usepackage{multirow}

\usepackage{orcidlink}

\newcommand{\keywords}[1]{\par\addvspace\baselineskip
\noindent\keywordname\enspace\ignorespaces#1}

\newcommand\extrafootertext[1]{%
    \bgroup
    \renewcommand\thefootnote{\fnsymbol{footnote}}%
    \renewcommand\thempfootnote{\fnsymbol{mpfootnote}}%
    \footnotetext[0]{#1}%
    \egroup
}

\urldef{\mailsa}\path|{mkunes,zhanzlic,jmatouse}@ntis.zcu.cz|

\begin{document}

\title{An Exploration of ECAPA-TDNN and x-vector Speaker Representations in Zero-shot Multi-speaker TTS}

\titlerunning{ECAPA-TDNN and x-vector Speaker Representations in Zero-shot TTS}

\iftoggle{blindforreview}{
    \author{[BLIND FOR REVIEW]}

    \authorrunning{[BLIND FOR REVIEW]}
    \institute{[BLIND FOR REVIEW]}
}{

    \author{Marie Kune\v{s}ov\'a\orcidlink{0000-0002-7187-8481} \and
    Zden{\v e}k Hanzl{\'i\v c}ek\orcidlink{0000-0002-4001-9289} \and
    Jind\v{r}ich Matou\v{s}ek\orcidlink{0000-0002-7408-7730}
    }

    
    \authorrunning{Marie Kune\v{s}ov\'a et al.}
    
    \institute{NTIS Research Centre and Department of Cybernetics,\\
        Faculty of Applied Sciences, University of West Bohemia in Pilsen, Czechia\\
        \mailsa\\
    }
    
    \index{Kune\v{s}ov\'a, Marie}
    \index{Hanzl{\'i\v c}ek, Zden{\v e}k}
    \index{Matou\v{s}ek, Jind\v{r}ich}
}

\toctitle{} \tocauthor{}

\maketitle

\begin{abstract}

Zero-shot multi-speaker text-to-speech (TTS) systems rely on speaker embeddings to synthesize speech in the voice of an unseen speaker, using only a short reference utterance. While many speaker embeddings have been developed for speaker recognition, their relative effectiveness in zero-shot TTS remains underexplored. In this work, we employ a YourTTS-based TTS system to compare three different speaker encoders -- YourTTS’s original H/ASP encoder, x-vector embeddings, and ECAPA-TDNN embeddings -- within an otherwise fixed zero-shot TTS framework. All models were trained on the same dataset of Czech read speech and evaluated on 24 out-of-domain target speakers using both subjective and objective methods. The subjective evaluation was conducted via a listening test focused on speaker similarity, while the objective evaluation measured cosine distances between speaker embeddings extracted from synthesized and real utterances. Across both evaluations, the original H/ASP encoder consistently outperformed the alternatives, with ECAPA-TDNN showing better results than x-vectors. These findings suggest that, despite the popularity of ECAPA-TDNN in speaker recognition, it does not necessarily offer improvements for speaker similarity in zero-shot TTS in this configuration. Our study highlights the importance of empirical evaluation when reusing speaker recognition embeddings in TTS and provides a framework for additional future comparisons.

\end{abstract}

\section{Introduction}

In recent years, zero-shot multi-speaker text-to-speech (TTS) has emerged as a rapidly advancing subfield of neural TTS. This approach allows a TTS system to synthesize speech in the voice of an unseen speaker, using only a short reference utterance of that speaker at inference time. A key component in such systems is the speaker encoder -- a model which extracts a speaker representation, typically in the form of a speaker embedding vector, from the reference audio. This representation can then be used to condition the TTS model and synthesize the voice of the target speaker. 
\extrafootertext{This preprint is a pre-peer-review version of the manuscript and has not undergone any post-submission improvements or corrections. The Version of Record of this contribution is expected to be published in the proceedings of the International Conference on Text, Speech, and Dialogue (TSD 2025).}

The choice of speaker encoder, along with the number of distinct speakers in the training data, serves an important role in determining the quality and speaker similarity of the resulting synthesized speech.

Most speaker encoders used in zero-shot TTS were originally created for speaker recognition, where the goal is to distinguish between speakers rather than to synthesize speech in their voice. Over the past few years, a wide variety of neural speaker embeddings have been proposed for speaker recognition tasks.

Examples include x-vectors \cite{Snyder2018xvector} and ECAPA-TDNN \cite{desplanques20_interspeech}, TitaNet \cite{koluguri2022titanet}, Deep Speaker \cite{li2017deepspeaker}, or various ``d-vectors'' (deep embedding vectors) \cite{variani2014,Wan2018_dvec}. More recently, self-supervised speech representation models such as WavLM~\cite{Chen2022wavlm} or wav2vec 2.0 and HuBERT \cite{yang21c_interspeech} have also been successfully used for speaker recognition.

Of this range of options, we took a particular interest in ECAPA-TDNN speaker embeddings. ECAPA-TDNN (Emphasized Channel Attention, Propagation and Aggregation – Time Delay Neural Network), like the closely related and still very popular x-vectors, builds upon the Time Delay Neural Network (TDNN) architecture. 
It is generally regarded as one of the top speaker encoders, having demonstrated state-of-the-art performance on speaker recognition and related tasks \cite{desplanques20_interspeech,dawalatabad21_interspeech,loweimi24_interspeech}.

Given its effectiveness, ECAPA-TDNN has begun to appear in TTS-related research as well. Recently, Jeong et al. \cite{Jeong2025} used an ECAPA-TDNN speaker encoder to support speaker-dependent duration prediction in a TTS model. Gusev et al. \cite{gusev24_interspeech} incorporated ECAPA-TDNN embeddings to compute a speaker loss in a zero-shot voice conversion system, and Li et al. \cite{Li2024SponTTS} employed ECAPA-TDNN-based embeddings to condition a HiFi-GAN decoder. These studies indicate growing interest in using speaker recognition models in TTS applications. However, they typically focus on system design rather than explicitly comparing speaker embeddings.

Several works have attempted a more direct comparison of different speaker embedding types in TTS systems. Xue et al.~\cite{Xue2022} compared multiple speaker encoders, including x-vectors and ECAPA-TDNN, within a FastSpeech2-based zero-shot TTS system. Cooper et al.~\cite{Cooper2020} explored learnable dictionary encoding-based speaker embeddings \cite{cai18_odyssey} and evaluated them against x-vectors in a Tacotron-based framework. Earlier, Doddipatla et al. \cite{doddipatla17_interspeech} performed a comparison of classic i-vectors \cite{Dehak2011ivec} and DNN-based d-vectors for speaker adaptation in TTS. Nevertheless, this area of comparative research remains underexplored -- especially in terms of isolating the effect of the speaker encoder while keeping the rest of the TTS system fixed.

In this paper, we aim to address this gap through a focused experimental study. We use YourTTS~\cite{casanova22yourtts}, a zero-shot multi-speaker TTS system, to evaluate and compare three types of speaker embeddings: the original speaker embeddings from the baseline system, x-vector embeddings, and ECAPA-TDNN embeddings. Our goal is not to propose a new architecture or embedding method, but to assess whether a different embedding extractor could improve speaker similarity in a plug-and-play manner.

This paper is organized as follows. In Section \ref{sec:TTS_system}, we describe the TTS system, the training and target speaker data, and the three speaker encoders. In Section \ref{sec:listening_test}, we present the design and results of a listening test created to compare the three TTS models, while Section \ref{sec:objective_eval} is dedicated to an objective evaluation, where we directly compare speaker embeddings of real and synthesized utterances. Finally, the last section summarizes our findings and presents our conclusions.

\section{TTS system}
\label{sec:TTS_system}

For our experiments, we used the YourTTS~\cite{casanova22yourtts} speech synthesis system, using a fork of the Coqui-ai/TTS\footnote{Original framework: \url{https://github.com/coqui-ai/TTS}; our baseline: 
\iftoggle{blindforreview}{[redacted for review]}{\url{https://github.com/jmaty/Coqui-TTS}}}
framework (throughout this paper, this system will be simply referred to as ``YourTTS''). 

YourTTS uses a VITS \cite{Kim2021vits} model with a modified approach that allows for zero-shot multi-speaker speech synthesis. The system uses a speaker encoder based on ``H/ASP'' \cite{heo2020clovabaselinevoxcelebspeaker} to obtain a speaker embedding from each utterance in the training data, which is used to condition the TTS model. By learning the speaker embedding space, the model is then able to synthesize speech from unseen speakers during inference, simply by being given the speaker embedding of a reference utterance.

The TTS system also employs an optional Speaker Consistency Loss (SCL) \cite{xin21_interspeech} during training. This is done by extracting speaker embeddings from the generated audio and comparing them to the embeddings of the original utterances using cosine distance. This process is meant to improve the voice similarity between the generated synthetic speech and the target speaker, but it requires that the speaker encoder be incorporated into the TTS system. Without SCL, it is possible to extract the embeddings for the training data externally and pass them to the TTS system in a separate input file at the beginning of the training process.

In this paper, our baseline system was trained with SCL, so the alternate options -- using different speaker encoders -- will employ it as well.

\subsection{Training Data}
\label{sec:train_data}

\iftoggle{blindforreview}{%
For the training of the TTS model, we used an in-house dataset 
}{%
For the training of the TTS model, we used an in-house dataset (``SPT-MGW'') 
}%
of Czech read speech recorded by 1116 speakers, with 150--174 short utterances (cca 2--10 seconds per utterance, $\sim$15 minutes in total) per speaker, sampled at 24 kHz. Approximately one quarter of the sentences in the dataset are identical across all speakers, while the rest of the texts had been randomly assigned from a larger list. 

This dataset was selected due to its large number of speakers, which should ensure good capability for synthesizing unseen voices. However, as the dataset had originally been intended for the training of robust automatic speech recognition (ASR) and was recorded by non-professional speakers in home environments, the acoustic quality and phonetic richness are somewhat lower than usual TTS datasets. 

Of the 1116 speakers in the dataset, we only used 1062 speakers for training, the remaining 54 speakers being excluded due to high levels of noise in the recordings. 

\subsection{Target speaker data}
\label{sec:target_speaker_data}

The reference utterances for the target speakers were sourced from radio broadcasts. We have access to a large collection of manually-labeled Czech radio broadcast data, spanning a period of more than a decade and containing utterances by thousands of different speakers. While the dataset's original annotations are not sufficiently accurate for it to be used for training (hence our choice of a different training dataset), they include the speaker identities, making it a useful source of target speaker references. 

For the purposes of these experiments, we selected 24 speakers from these data, 12 men and 12 women, ranging from well-known personalities with highly recognizable voices (prominent politicians and actors) to less publicly known individuals (radio moderators, various interviewees). As far as we are aware, none of these speakers were present in the training dataset.

For embedding extraction, we selected suitable radio recordings for each speaker, initially aiming to obtain approximately 2 minutes of speech per speaker. During the selection process, the recordings were manually checked to ensure consistent acoustic conditions with no background noise, particularly avoiding any instances of other speakers talking. Where possible, we also tried to select the recordings for each speaker from a relatively narrow time range (ideally from the same broadcast) to avoid any changes in a speaker's voice that may occur over time. 

From the selected recordings, we first set one utterance per speaker aside to be used as a reference in the listening test. The remaining files from each speaker were resampled to 16\,kHz (as required by the embedding extractors), concatenated in a random order, and cut to a uniform length of 30 seconds. These 30-second samples were then used as target speaker references for speech synthesis by passing them to the TTS system at inference time in lieu of a speaker ID. The 30-second duration was eventually chosen because the baseline speaker encoder uses, at most, slightly less than 30 seconds of audio to extract an embedding, regardless of the available duration (see the next section for details). Limiting the reference to this length thus ensures a fairer comparison between different encoders.

\subsection{Speaker encoders}

We replaced the TTS system's default speaker encoder (which, as mentioned, is based on the ``H/ASP'' approach in~\cite{heo2020clovabaselinevoxcelebspeaker}) with code from the SpeechBrain toolkit~\cite{ravanelli2021speechbrain}, using either the x-vector or the ECAPA-TDNN models available in SpeechBrain. Then, we trained the TTS system for multi-speaker speech synthesis using our training dataset.

In total, we trained three TTS models:

\begin{enumerate}
    \item ``H/ASP TTS'' -- baseline system using the original pretrained speaker encoder used in YourTTS (embedding dimension: 512, trained on the VoxCeleb2 dataset \cite{chung18b_interspeech}) 
    \item ``x-vector TTS'' -- using a SpeechBrain speaker encoder, with a pretrained x-vector model\footnote{\url{https://huggingface.co/speechbrain/spkrec-xvect-voxceleb}} (embedding dimension: 512, trained on the VoxCeleb 1 \cite{nagrani17_interspeech} and VoxCeleb2 datasets)
    \item ``ECAPA-TDNN TTS'' -- using a SpeechBrain speaker encoder, with a pretrained ECAPA-TDNN model\footnote{\url{https://github.com/speechbrain/speechbrain/tree/develop/recipes/VoxCeleb/SpeakerRec}
    } (embedding dimension: 192, trained on the VoxCeleb 1 and 2 datasets)
\end{enumerate}

Each model was trained for 295 epochs (approximately 1.6 million steps). Table \ref{tab:model_settings} shows the most important model settings and parameters.

\begin{table}[ht]
    \caption{Model settings during training}
    \label{tab:model_settings}
    \centering
    \begin{tabular}{|l|l|}
    \hline
        Setting & Value \\
    \hline
        \verb|use_speaker_encoder_as_loss| (SCL)& True \\
        \verb|d_vector_dim| (embedding dimension) & baseline, x-vector: 512 \\
        & ECAPA-TDNN: 192 \\
        \verb|epochs| & 295 (all models)\\
        \verb|hop_length| & 256 \\
        \verb|sample_rate| & 24000 \\
        \verb|mixed_precision| & False \\
    \hline
    \end{tabular}
    
\end{table}

It should also be noted that the individual speaker encoders differ in how they process the 30-second-long reference utterances:

\begin{itemize}
    \item The original H/ASP encoder, as implemented in YourTTS, does not extract embeddings from the entire audio file. Instead, the TTS system selects several short segments, equally spaced throughout the reference audio file, extracts an embedding from each, and then calculates the mean of these embeddings. 
    By default, it uses 10 segments, each with a length of 250 $\times$ \texttt{hop\_length} audio samples, which, in our case, equals $\sim$2.67 seconds per segment, for a total of $\sim$26.7 seconds of processed audio. The same process is also used during training.  
    
    \item By contrast, since we completely bypassed YourTTS's \texttt{compute\_embedding} function with our modifications, the alternate speaker encoders do not split the input audio in that manner. Instead, the SpeechBrain encoder processes the entire audio file at once. The duration of the input file does not matter, thanks to the statistics pooling present in both the x-vector and ECAPA-TDNN architectures. 

\end{itemize}

\section{Listening test}
\label{sec:listening_test}

In order to compare the three speaker encoders, we performed both subjective and objective evaluations. The subjective evaluation took the form of a listening test with 24 sets of samples -- one for each test speaker. The listeners had to rate the similarity between a reference utterance of a target speaker and three samples synthesized by the TTS models, on a 0--100 scale, similar to MUSHRA \cite{ITU_MUSHRA}. 

The synthesized sentence was identical for all speakers and models:
\iftoggle{blindforreview}{%
``Tuto v\v{e}tu nejsp\'{i}\v{s} nikdy ne\v{r}eknu, ale byla vytvo\v{r}ena pomoc\'{i} synt\'{e}zy \v{r}e\v{c}i na [REDACTED FOR REVIEW].'' English translation: ``I will probably never say this sentence, but it was created using speech synthesis at [REDACTED FOR REVIEW]''.
}{%
``Tuto v\v{e}tu nejsp\'{i}\v{s} nikdy ne\v{r}eknu, ale byla vytvo\v{r}ena pomoc\'{i} synt\'{e}zy \v{r}e\v{c}i na kated\v{r}e kybernetiky.'' English translation: ``I will probably never say this sentence, but it was created using speech synthesis at the Department of Cybernetics.''
} 

The reference utterances for the listening set were the ones that we had originally set aside for this purpose, as mentioned in Section \ref{sec:target_speaker_data}, in some cases cut down to a more suitable length. The durations of these utterances ranged from 3 to 7 seconds.

A total of 23 listeners participated in the listening test, all of whom were native Czech speakers.
The listeners were recruited among our colleagues and their family members, and also included several students. Of the 23 listeners, 12 had extensive prior experience with listening tests, while 11 were non-experts with little to no experience with listening tests or TTS in general. All listeners evaluated the same speech samples.

Given the imperfect nature of the training dataset (as mentioned in Section \ref{sec:train_data}), our listening test only focused on speaker similarity. The listeners were instructed to only rate the similarity between speakers' voices and to ignore the quality of the synthesized speech. They were also given the following recommended rating guidelines (translated here from the original Czech):

\begin{itemize}
    \item[100] \dots ``Definitely identical voice''
    \item[$\sim$75] \dots ``Very similar voice (sounds almost the same), could be the same person''
    \item[$\sim$50] \dots ``The voices are quite similar, but it's a 50/50 chance that it's the same person''
    \item[$\sim$25] \dots ``The voices are somewhat similar, but don't sound like the same person''
    \item[0] \dots ``The voices are not at all similar''
\end{itemize}


\begin{figure}[ht]
    \centering
    \includegraphics[width=0.75\linewidth]{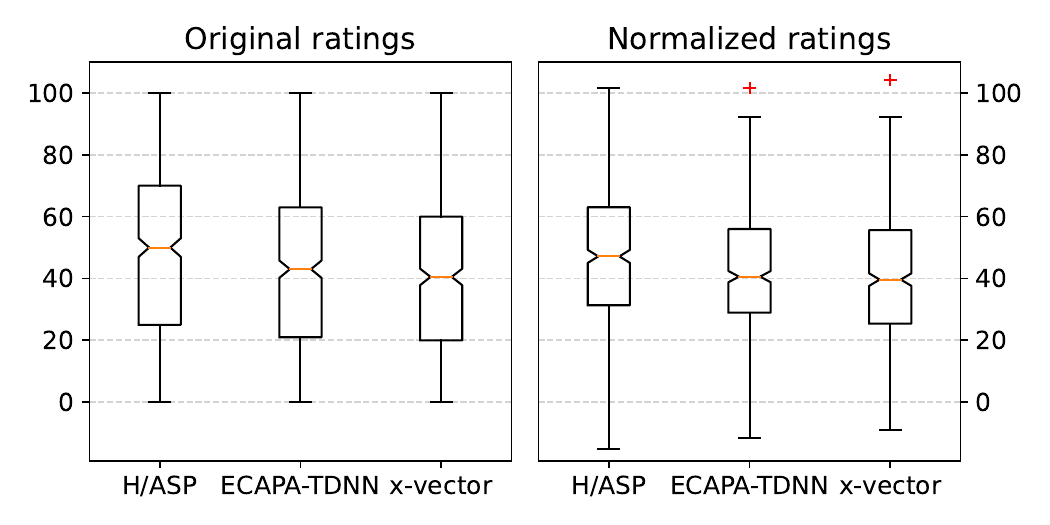}
    \caption{A visualization of the listening test results, before and after normalization.}
    \label{fig:listening_test}
\end{figure}

The results of the listening test are shown in Table~\ref{tab:listening_test} and Figure~\ref{fig:listening_test}. To aid our evaluation and compensate for individual listener biases, the answers from each listener were normalized to the same mean and variance, and we show the statistics for both the normalized ratings and the original raw values given by the listeners. (Note that due to the normalization, some values now exceed the original range of 0--100, as seen in Fig.~\ref{fig:listening_test}.)

\begin{table}[ht]
    \centering
    \caption{Results of the listening test as means, standard deviations, and medians of all individual ratings from 23 listeners and 24 questions. A higher value indicates a higher degree of similarity to the target speaker.}
    \label{tab:listening_test}
    \begin{tabular}{|l|c|c|c|c|}
        \hline
        \multirow{2}{*}{model} & \multicolumn{2}{c|}{raw ratings} & \multicolumn{2}{c|}{normalized ratings}\\
        \cline{2-5}
        & mean $\pm$ std & median & mean $\pm$ std & median\\
        \hline

        H/ASP TTS (baseline) & \textbf{47.28 $\pm$ 26.19} & \textbf{50.0} & \textbf{47.31 $\pm$ 20.50} & \textbf{47.14}\\
        ECAPA-TDNN TTS & 42.62 $\pm$ 26.52 & 43.0 & 42.61 $\pm$ 20.38 & 40.59 \\
        x-vector TTS & 40.98 $\pm$ 26.24 & 40.5 & 40.96 $\pm$ 20.85 & 39.63 \\
        \hline
    \end{tabular}
\end{table}

To evaluate the statistical significance of these results, we analyzed them using the Wilcoxon signed-rank test, with the Holm-Bonferroni correction for multiple comparisons. The test showed a statistically significant difference for all pairwise combinations of models: the baseline ``H/ASP TTS'' model was rated better than the other two TTS models ($p < 0.001$ in both cases), and ``ECAPA-TDNN TTS'' was rated better than ``x-vector TTS'' ($p = 0.03$).

\section{Objective evaluation}
\label{sec:objective_eval}

In addition to the subjective listening test, we also performed a more objective evaluation of speaker similarity, using speaker recognition techniques.

In the field of speaker recognition, the most common approach to verifying if two utterances were produced by the same speaker is to extract speaker embeddings from each utterance and then compare them, typically by calculating their cosine distance~\cite{desplanques20_interspeech,koluguri2022titanet}. Utterances from the same speaker should result in a low cosine distance, while for different speakers, the distance should be high, though the exact values may depend on the process used to extract the embeddings. In this paper, we employ the same approach to express the level of similarity between the synthesized samples and the real utterances of the target speakers.

First, to obtain more robust results, we synthesized 20 additional sentences for each target speaker and model combination (for a total of 21, including those used in the listening test). Then, we extracted speaker embeddings from each synthesized utterance, the references used in the listening test, and 15 other real utterances of each speaker, and measured their similarity using cosine distance. 

The 15 representative utterances from each speaker were selected from the same dataset and time range as the previously used data, but did not include the reference utterance from the listening test. The durations of these utterances ranged from 3 to 30 seconds.

Since the results of this evaluation may depend on the specific method used to extract the speaker embeddings, we performed the comparison four times, using different speaker embedding extractors: the previously-used ECAPA-TDNN and x-vector models, NVIDIA's \texttt{TitaNet-large} \cite{koluguri2022titanet} model (via the NeMo toolkit \cite{kuchaiev2019nemotoolkit}), 
and the \texttt{Resemblyzer} Python package\footnote{\url{https://pypi.org/project/Resemblyzer/}}. For the first three of these extractors, the extracted embeddings were also mean and L2 normalized (the embeddings from all audio files, both natural and synthesized, were normalized together). However, this normalization was not applied to Resemblyzer embeddings, as the Resemblyzer package already performs L2 normalization as part of its embedding extraction process. 

\begin{figure}[ht]

\includegraphics[width=\textwidth]{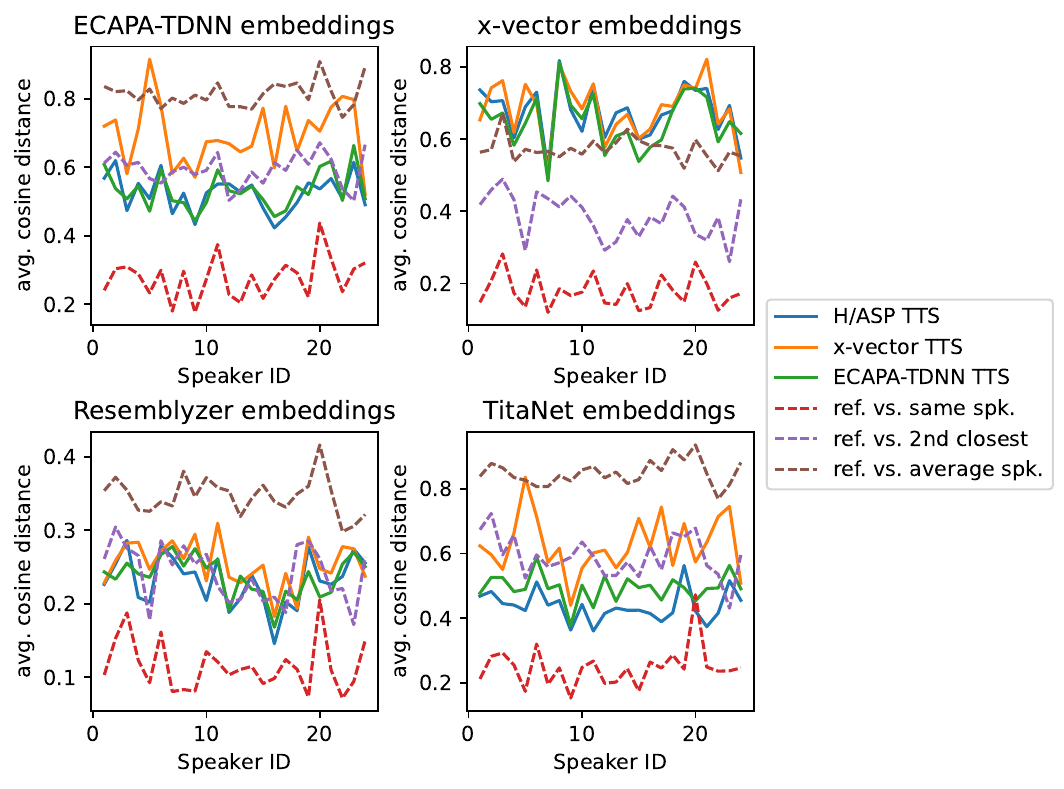} 

\caption{Average cosine distances between speaker embeddings extracted using different models and representative utterances of the target speakers. Each graph represents a separate evaluation using a different embedding extractor.} 

\label{fig:graphs_embed_dists}
\end{figure}

Figure~\ref{fig:graphs_embed_dists} shows the results of this comparison. Each graph plots the average cosine distances between the 21 synthesized sentences and the 15 natural utterances from the same speaker, calculated from embeddings extracted using a specific model. For comparison, the graphs also show, with dashed lines, three sets of distances for the reference utterances from the listening test, namely: 
\begin{itemize}
    \item the average distance to other recordings of the same speaker, which can naturally be expected to be low,
    \item the average distance to all recordings of different speakers, which can be expected to be high,
    \item and the average distance to the recordings of the second most similar speaker (of the 24 speakers used in the listening test). 
\end{itemize}

The distances shown in Figure~\ref{fig:graphs_embed_dists} were calculated as:
\begingroup
\allowdisplaybreaks 
\begin{align*}
    & \mathrm{avg\_dist\_ref}(r_i,S_i) = \frac{1}{K}\sum_{k} \mathrm{emb\_dist} (r_i,s_{i,k})\\
    & \mathrm{avg\_dist\_tts}(U_i,S_i) = \frac{1}{K}\sum_{k} \frac{1}{M}\sum_{m} \mathrm{emb\_dist} (u_{i,m},s_{i,k})\\
    \text{H/ASP TTS:}\quad & \mathrm{dist}_{\mathrm{H/ASP}}(i) = \mathrm{avg\_dist\_tts} (U_{i,\mathrm{H/ASP}},S_i) \\
    \text{x-vector TTS:}\quad & \mathrm{dist}_{\mathrm{xvec}}(i) = \mathrm{avg\_dist\_tts} (U_{i,\mathrm{xvec}},S_i)\\
    \text{ECAPA-TDNN TTS:}\quad & \mathrm{dist}_{\mathrm{ECAPA}}(i) = \mathrm{avg\_dist\_tts} (U_{i,\mathrm{ECAPA}},S_i) \\
    \text{ref. vs. same spk.:}\quad & \mathrm{dist}_\mathrm{same}(i) = \mathrm{avg\_dist\_ref} (r_i,S_i)\\
    \text{ref. vs. 2nd closest:}\quad & \mathrm{dist}_\mathrm{2nd\_closest}(i) = \min_{j\neq i} \mathrm{avg\_dist\_ref} (r_i,S_j)  \\
    \text{ref. vs. average spk.:}\quad & \mathrm{dist}_\mathrm{avg}(i) = \frac{1}{N-1}\sum_{j\neq i} \mathrm{avg\_dist\_ref} (r_i,S_j),
\end{align*}%
\endgroup
where $s_{i,1} \dots s_{i,K}$ are the $K=15$ representative natural utterances for speaker $S_i$; $U_i = \{u_{i,1} \dots u_{i,m}$\} is a set of $M=21$ synthesized utterances for target speaker $S_i$, synthesized using H/ASP ($U_{i,\mathrm{H/ASP}}$), x-vector ($U_{i,\mathrm{xvec}}$), or ECAPA-TDNN  ($U_{i,\mathrm{ECAPA}}$) speaker encoders; $r_i$ is the reference utterance used in the listening test for speaker $S_i$; ``$\mathrm{emb\_dist}(x,y)$'' denotes the cosine distance between embeddings extracted from utterances $x$ and $y$, and $N=24$ is the total number of speakers. 

As seen in Figure~\ref{fig:graphs_embed_dists}, the results for different embedding models are not fully consistent, but there are some clear trends: 
In two out of four cases (using embeddings extracted with ECAPA-TDNN and TitaNet), the utterances synthesized with the ``x-vector TTS'' model consistently have a higher distance from the target speaker than the other two TTS models, often by a large margin. According to the TitaNet embeddings, the baseline H/ASP model is also consistently better than the ECAPA-TDNN model.
With the other two types of embeddings (usng Resemblyzer and x-vectors), the plotted lines for the three TTS models are very close to each other, and their relative placements fluctuate. However, even here we can safely say that overall, neither of the alternate models has managed to clearly surpass the baseline.

These observations are also supported by Table~\ref{tab:obj_results}, which shows the means and standard deviations of the same calculated distances across the 24 target speakers. Note that since each column represents a separate evaluation using a different embedding extractor, the values in different columns cannot be compared with each other.

\begin{table}[ht]
    \centering
    \caption{Cosine distance between speaker embeddings extracted using different models, as means and standard deviations across the 24 target speakers. A lower value indicates a higher degree of similarity.}
    \label{tab:obj_results}
    \begin{tabular}{|c|c|c|c|c|}
    \hline
    & \multicolumn{4}{c|}{speaker embedding extractor} \\
    \cline{2-5}
    &  ECAPA-TDNN & x-vector & Resemblyzer & TitaNet\\
    \hline 

    \textbf{H/ASP TTS (baseline)} & \textbf{0.524 $\pm$ 0.052} & 0.674 $\pm$ 0.070 & \textbf{0.438 $\pm$ 0.046} & \textbf{0.231 $\pm$ 0.034}\\
    ECAPA-TDNN TTS & 0.533 $\pm$ 0.054 & \textbf{0.649 $\pm$ 0.074} & 0.494 $\pm$ 0.042 & 0.237 $\pm$ 0.027\\
    x-vector TTS & 0.696 $\pm$ 0.090 & 0.681 $\pm$ 0.081 & 0.627 $\pm$ 0.086 & 0.254 $\pm$ 0.030\\
    \hline
    ref. vs. same spk. & 0.277 $\pm$ 0.059 & 0.178 $\pm$ 0.044 & 0.248 $\pm$ 0.061 & 0.116 $\pm$ 0.034\\
    ref. vs. 2nd closest & 0.593 $\pm$ 0.045 & 0.386 $\pm$ 0.060 & 0.591 $\pm$ 0.063 & 0.241 $\pm$ 0.037\\
    ref. vs. average spk. & 0.814 $\pm$ 0.037 & 0.573 $\pm$ 0.032 & 0.849 $\pm$ 0.037 & 0.346 $\pm$ 0.025\\
    
    \hline
    \end{tabular}

\end{table}

Finally, as with the listening test, we also performed a statistical analysis of these results, again using pairwise comparisons and the Wilcoxon signed-rank test with a Holm-Bonferroni correction for multiple comparisons. This was based on the average embedding distances per speaker and embedding extractor, across all speakers and extractors (i.e., using $24 \times 4 = 96$ values per TTS model). The results of the statistical test confirm that the ``ECAPA-TDNN TTS'' and ``H/ASP TTS'' models are both significantly better than ``x-vector TTS'' ($p < 0.001$ in both cases), and also suggest that ``H/ASP TTS'' is better than ``ECAPA-TDNN TTS'' ($p = 0.02$). This matches the results of the listening test, which were presented in Section \ref{sec:listening_test}.

\section{Conclusions}

In this paper, we have explored the use of two alternative speaker encoder models, x-vector and ECAPA-TDNN, for speaker representation in a zero-shot multi-speaker TTS system based on YourTTS. We evaluated the speaker similarity of utterances synthesized with the three embedding types using both subjective and objective metrics, leading to very similar results. Our experimental findings suggest that while ECAPA-TDNN appears to be more suitable for this task than x-vector embeddings, the original H/ASP encoder still remains the most effective in this context. Nevertheless, we view this outcome not as a disappointment but as a valuable insight into the comparative behavior of speaker embeddings in zero-shot TTS.

By isolating the effect of the speaker encoder within a common TTS framework, this study contributes to a better understanding of how speaker embeddings impact performance in zero-shot synthesis. We believe these findings can inform future research and development of more robust and generalizable TTS systems.

Although this experiment did not yield improvements over the baseline,
we would still like to explore other alternative speaker encoders, such as the TitaNet-large model we employed in Section \ref{sec:objective_eval}. Outside of the present work, we have already experimented with the use of TitaNet-large as an external embedding extractor for TTS -- without employing speaker consistency loss -- with promising results. In the future, we would like to incorporate this model into the TTS system itself, as we did here with the SpeechBrain encoder, so that it can be properly compared with the other approaches.

\iftoggle{blindforreview}{
}{
    \begin{credits}
    \subsubsection{\ackname} 

    This work was created with the support of the project ``R\&D of Technologies for Advanced Digitization in the Pilsen Metropolitan Area (DigiTech)'' No.: CZ.02.01.01/00/\allowbreak 23\_021/0008436 co-financed by the European Union.    
    Computational resources were provided by the e-INFRA CZ project (ID:90254), supported by the Ministry of Education, Youth and Sports of the Czech Republic.
    
    \subsubsection{\discintname}
    The authors have no competing interests to declare that are relevant to the content of this article.

    \end{credits}
}
%
%

\bibliographystyle{splncs04}
\bibliography{bibliography}

\begin{thebibliography}{10}
\providecommand{\url}[1]{\texttt{#1}}
\providecommand{\urlprefix}{URL }
\providecommand{\doi}[1]{https://doi.org/#1}

\bibitem{cai18_odyssey}
Cai, W., Chen, J., Li, M.: Exploring the encoding layer and loss function in end-to-end speaker and language recognition system. In: The Speaker and Language Recognition Workshop (Odyssey 2018). pp. 74--81 (2018). \doi{10.21437/Odyssey.2018-11}

\bibitem{casanova22yourtts}
Casanova, E., Weber, J., Shulby, C.D., Candido~Jr., A., G{\"o}lge, E., Ponti, M.A.: {YourTTS}: Towards zero-shot multi-speaker {TTS} and zero-shot voice conversion for everyone. In: Proceedings of the 39th International Conference on Machine Learning. Proceedings of Machine Learning Research, vol.~162, pp. 2709--2720. PMLR (2022), \url{https://proceedings.mlr.press/v162/casanova22a.html}

\bibitem{Chen2022wavlm}
Chen, S., et~al.: {WavLM:} large-scale self-supervised pre-training for full stack speech processing. IEEE Journal of Selected Topics in Signal Processing  \textbf{16}(6),  1505--1518 (2022). \doi{10.1109/JSTSP.2022.3188113}

\bibitem{chung18b_interspeech}
Chung, J.S., Nagrani, A., Zisserman, A.: {VoxCeleb2:} deep speaker recognition. In: Interspeech 2018. pp. 1086--1090 (2018). \doi{10.21437/Interspeech.2018-1929}

\bibitem{Cooper2020}
Cooper, E., Lai, C.I., Yasuda, Y., Fang, F., Wang, X., Chen, N., Yamagishi, J.: Zero-shot multi-speaker text-to-speech with state-of-the-art neural speaker embeddings. In: ICASSP 2020 - 2020 IEEE International Conference on Acoustics, Speech and Signal Processing (ICASSP). pp. 6184--6188 (2020). \doi{10.1109/ICASSP40776.2020.9054535}

\bibitem{dawalatabad21_interspeech}
Dawalatabad, N., Ravanelli, M., Grondin, F., Thienpondt, J., Desplanques, B., Na, H.: {ECAPA-TDNN} embeddings for speaker diarization. In: Interspeech 2021. pp. 3560--3564 (2021). \doi{10.21437/Interspeech.2021-941}

\bibitem{Dehak2011ivec}
Dehak, N., Kenny, P.J., Dehak, R., Dumouchel, P., Ouellet, P.: Front-end factor analysis for speaker verification. IEEE Transactions on Audio, Speech, and Language Processing  \textbf{19}(4),  788--798 (2011). \doi{10.1109/TASL.2010.2064307}

\bibitem{desplanques20_interspeech}
Desplanques, B., Thienpondt, J., Demuynck, K.: {ECAPA-TDNN}: Emphasized channel attention, propagation and aggregation in {TDNN} based speaker verification. In: Interspeech 2020. pp. 3830--3834 (2020). \doi{10.21437/Interspeech.2020-2650}

\bibitem{doddipatla17_interspeech}
Doddipatla, R., Braunschweiler, N., Maia, R.: Speaker adaptation in {DNN}-based speech synthesis using d-vectors. In: Interspeech 2017. pp. 3404--3408 (2017). \doi{10.21437/Interspeech.2017-1038}

\bibitem{gusev24_interspeech}
Gusev, A., Avdeeva, A.: Improvement speaker similarity for zero-shot any-to-any voice conversion of whispered and regular speech. In: Interspeech 2024. pp. 2735--2739 (2024). \doi{10.21437/Interspeech.2024-2091}

\bibitem{heo2020clovabaselinevoxcelebspeaker}
Heo, H.S., Lee, B.J., Huh, J., Chung, J.S.: Clova baseline system for the {VoxCeleb Speaker Recognition Challenge} 2020. arXiv preprint arXiv:2009.14153  (2020). \doi{10.48550/arXiv.2009.14153}

\bibitem{ITU_MUSHRA}
{ITU-R Recommendation BS.1534-3}: Method for the subjective assessment of intermediate quality level of audio systems. Tech. rep., International Telecommunication Union (2015)

\bibitem{Jeong2025}
Jeong, M., Kim, M., Kim, S., Kim, N.S.: {Evidential-TTS:} high fidelity zero-shot text-to-speech using evidential deep learning. In: ICASSP 2025 - 2025 IEEE International Conference on Acoustics, Speech and Signal Processing (ICASSP). pp.~1--5 (2025). \doi{10.1109/ICASSP49660.2025.10889279}

\bibitem{Kim2021vits}
Kim, J., Kong, J., Son, J.: Conditional variational autoencoder with adversarial learning for end-to-end text-to-speech. In: Proceedings of the 38th International Conference on Machine Learning. vol.~139, pp. 5530--5540. PMLR (2021), \url{https://proceedings.mlr.press/v139/kim21f.html}

\bibitem{koluguri2022titanet}
Koluguri, N.R., Park, T., Ginsburg, B.: {TitaNet:} neural model for speaker representation with {1D} depth-wise separable convolutions and global context. In: ICASSP 2022 - 2022 IEEE International Conference on Acoustics, Speech and Signal Processing (ICASSP). pp. 8102--8106 (2022). \doi{10.1109/ICASSP43922.2022.9746806}

\bibitem{kuchaiev2019nemotoolkit}
Kuchaiev, O., et~al.: {NeMo}: a toolkit for building {AI} applications using neural modules. arXiv preprint arXiv:1909.09577  (2019). \doi{10.48550/arXiv.1909.09577}

\bibitem{li2017deepspeaker}
Li, C., Ma, X., Jiang, B., Li, X., Zhang, X., Liu, X., Cao, Y., Kannan, A., Zhu, Z.: {Deep Speaker}: an end-to-end neural speaker embedding system (2017). \doi{10.48550/arXiv.1705.02304}

\bibitem{Li2024SponTTS}
Li, H., Zhu, X., Xue, L., Song, Y., Chen, Y., Xie, L.: {SponTTS:} modeling and transferring spontaneous style for {TTS}. In: ICASSP 2024 - 2024 IEEE International Conference on Acoustics, Speech and Signal Processing (ICASSP). pp. 12171--12175 (2024). \doi{10.1109/ICASSP48485.2024.10445828}

\bibitem{loweimi24_interspeech}
Loweimi, E., Qian, M., Knill, K., Gales, M.: On the usefulness of speaker embeddings for speaker retrieval in the wild: A comparative study of x-vector and {ECAPA-TDNN} models. In: Interspeech 2024. pp. 3774--3778 (2024). \doi{10.21437/Interspeech.2024-161}

\bibitem{nagrani17_interspeech}
Nagrani, A., Chung, J.S., Zisserman, A.: {VoxCeleb:} a large-scale speaker identification dataset. In: Interspeech 2017. pp. 2616--2620 (2017). \doi{10.21437/Interspeech.2017-950}

\bibitem{ravanelli2021speechbrain}
Ravanelli, M., et~al.: {SpeechBrain}: A general-purpose speech toolkit. arXiv preprint arXiv:2106.04624  (2021). \doi{10.48550/arXiv.2106.04624}

\bibitem{Snyder2018xvector}
Snyder, D., Garcia-Romero, D., Sell, G., Povey, D., Khudanpur, S.: X-vectors: Robust {DNN} embeddings for speaker recognition. In: 2018 IEEE International Conference on Acoustics, Speech and Signal Processing (ICASSP). pp. 5329--5333 (2018). \doi{10.1109/ICASSP.2018.8461375}

\bibitem{variani2014}
Variani, E., Lei, X., McDermott, E., Lopez~Moreno, I., Gonzalez-Dominguez, J.: Deep neural networks for small footprint text-dependent speaker verification. In: 2014 IEEE International Conference on Acoustics, Speech and Signal Processing (ICASSP). pp. 4052--4056 (2014). \doi{10.1109/ICASSP.2014.6854363}

\bibitem{Wan2018_dvec}
Wan, L., Wang, Q., Papir, A., Lopez~Moreno, I.: Generalized end-to-end loss for speaker verification. In: 2018 IEEE International Conference on Acoustics, Speech and Signal Processing (ICASSP). pp. 4879--4883 (2018). \doi{10.1109/ICASSP.2018.8462665}

\bibitem{xin21_interspeech}
Xin, D., Saito, Y., Takamichi, S., Koriyama, T., Saruwatari, H.: Cross-lingual speaker adaptation using domain adaptation and speaker consistency loss for text-to-speech synthesis. In: Interspeech 2021. pp. 1614--1618 (2021). \doi{10.21437/Interspeech.2021-897}

\bibitem{Xue2022}
Xue, J., Deng, Y., Han, Y., Li, Y., Sun, J., Liang, J.: {ECAPA-TDNN} for multi-speaker text-to-speech synthesis. In: 2022 13th International Symposium on Chinese Spoken Language Processing (ISCSLP). pp. 230--234 (2022). \doi{10.1109/ISCSLP57327.2022.10037956}

\bibitem{yang21c_interspeech}
{Yang}, S., et~al.: {SUPERB: Speech Processing Universal PERformance Benchmark}. In: Interspeech 2021. pp. 1194--1198 (2021). \doi{10.21437/Interspeech.2021-1775}

\end{thebibliography}

\end{document}